# Negative refraction in Al:ZnO/ZnO metamaterial in the near-infrared

Gururaj V. Naik[1], Jingjing Liu[1], Alexander V. Kildishev[1], Vladimir M. Shalaev[1] and Alexandra Boltasseva[1,2,*]

[1]School of Electrical & Computer Engineering and Birck Nanotechnology Center, Purdue University

West Lafayette, IN 47907, USA.

[2]DTU Fotonik, Technical University of Denmark, Lyngby 2800, Denmark.

*Corresponding author: aeb@purdue.edu

Plasmonics[1-2] and metamaterials[3] have brought a paradigm shift in the scope and outlook of optics by enabling novel functionalities such as sub-diffraction imaging[4-5], cloaking[6] and optical magnetism[7]. Many of these functionalities have already been demonstrated as proof-of-principle experiments performed in the ultraviolet (UV), infrared (IR) or microwave frequency range[4-6,8]. Experimental demonstrations of similar devices in the visible and near-IR ranges would have a significantly larger impact owing to the technological importance of these frequency ranges. However, the major problem impeding the realization of efficient devices in these important spectral ranges is the optical losses associated with metallic parts in the metamaterial building blocks[9]. The development of new constituent materials for low-loss metamaterial-based devices is therefore required[10]. Previously, we reported that heavily doped oxide semiconductors like transparent conducting oxides could be good alternatives to metals for the realization of many classes of metamaterials, including metamaterials with hyperbolic dispersion[11-12] in the near-IR. Here, for the first time, we demonstrate a negatively refracting metamaterial formed by replacing metal with aluminum-doped zinc oxide (AZO). We show that the performance of this metamaterial device is significantly better than the conventional metal-based devices.

Optical metamaterials typically use nanostructured metals and dielectrics as their building blocks. The geometry of the unit cell and the optical properties of the constituent materials form the parameter space for the device design. In the near-IR and shorter wavelength ranges, the geometry is restricted by nanofabrication capabilities. Thus, the optical properties of the constituent materials form an important set of design parameters. The optical properties of materials are described by their complex-valued permittivities and permeabilities. The permeabilities of natural materials are unity in the optical frequencies. Hence, optical properties of these materials are described by their permittivities or dielectric functions ($\varepsilon$). The real part of the dielectric function signifies the electrical polarization response of the material, while the imaginary part describes the losses. When designing a low-loss, high-performance metamaterial, one can choose dielectric materials (for positive ($\varepsilon'$) contribution) that have negligible losses. However, metals that offer negative real permittivity are accompanied by large losses. Even noble metals like silver and gold, which are the metals with the highest DC conductivities, exhibit excessive losses at optical frequencies that seriously restrict the experimental realization of novel metamaterials devices in the optical frequency range. Furthermore, the real part of the permittivity for metals is very large in magnitude, which causes additional problems in the design of metamaterial devices. This is because many devices based on transformation optics[13] require the polarization responses of the metallic and dielectric constituents to be nearly balanced[14]. Balancing these responses requires that the magnitudes of the real permittivities of the metal and dielectric components be comparable. At optical frequencies, the real permittivity of dielectrics is on the order of 1 to 10 (e.g., dielectric permittivity of $Al_2O_3$ is about 3 in the optical range), while that of metals is about one to two orders higher (−115 for Ag at 1.5 µm). Additionally, there are other important drawbacks when working with conventional metals (Au and Ag): their optical properties are not tunable, and processing of these metals is not compatible with standard silicon nanofabrication technology. Such technological incompatibilities preclude the integration of plasmonics and metamaterials with nanoelectronics, which could bring about a wide range of new applications.

Semiconductors offer great tunability in their properties (e.g. by doping). However, conventional semiconductors are dielectrics in the optical range and, hence, their use in plasmonics and optical metamaterials has been limited to being dielectrics. Heavy doping can make semiconductors exhibit metal-like properties in the mid-IR[15], but the typical doping levels of about $10^{18-19}$ cm$^{-3}$ cannot make semiconductors exhibit metallic properties in the near-IR. Further, higher doping (~$10^{21}$ cm$^{-3}$) is required to make semiconductors metallic at optical frequencies. Such ultra-high doping in semiconductors poses significant challenges because of solid-solubility limits[16-17] and low dopant ionization efficiencies[18]. Oxide semiconductors allow ultra-high doping as a result of their higher solid-solubility limits[19], so these materials could exhibit metallic properties in the near-IR. Heavily doped zinc oxide, one of a family of transparent conducting oxides (TCOs), is a highly conducting material that has been used in applications such as liquid-crystal displays. Zinc oxide can be heavily doped (~$10^{21}$ cm$^{-3}$) by trivalent dopants such as aluminum and gallium. Such high levels of doping are not possible in conventional semiconductors such as silicon or GaAs because of low solid-solubility limits of dopants[16-17]. Ultra-high doping results in a large carrier concentration ($10^{20-21}$ cm$^{-3}$), which enables Drude metal-like optical properties in the near-IR. This can result in negative values of real permittivity for doped zinc oxide and other TCOs. While the carrier concentration is related to the metallic properties of TCO films, the carrier relaxation rate or Drude damping relates to the optical losses in TCO films[20]. Since losses are detrimental to the performance of the devices, Drude damping needs to be as low as possible in these materials. Previously, we studied various TCOs and found that aluminum-doped zinc oxide (AZO) exhibits the lowest losses, about five times lower than the loss in silver in the near-IR, owing to its small Drude-damping coefficient[10]. Here, we demonstrate a high-performance hyperbolic metamaterial (HMM) formed by AZO and ZnO as the metallic and dielectric components, respectively.

A uniaxial material that has different signs of real permittivity in different directions is a HMM[11-12]. Consequently, there are two possibilities for this to occur: the real permittivity parallel to the uniaxial crystal planes is positive ($\varepsilon_\parallel > 0$) and that perpendicular to the planes is negative ($\varepsilon_\perp < 0$), herein referred to as a Type-1 HMM; or $\varepsilon_\parallel < 0$ and $\varepsilon_\perp > 0$, which we label as a Type-2 HMM. It is only the Type-1 HMM that can exhibit negative refraction or hyperlensing[4,15,21-22]. Such applications are enabled by realizing hyperbolic dispersion in metal-dielectric composite structures. Sub-wavelength building blocks of metallic and dielectric components could exhibit the desired uniaxial anisotropy in the effective-medium limit[11,23]. Two types of design are used in HMMs: a planar design consisting of alternating layers of metal and dielectric[15,24], and metal nanowires embedded in a dielectric medium[21-22]. The planar design has many advantages over the nanowire structure because its fabrication is simple and robust. Moreover, it can be easily integrated with other components into more complex device geometries using standard, planar fabrication processes. However, achieving Type-1 hyperbolic dispersion in a planar structure is very challenging, especially in the near-IR and visible ranges. This is because metals that are typically used to build such a planar metamaterial have very large magnitude of real permittivity. In other words, conventional metals such as silver and gold are too metallic to be useful as building blocks for a hyperlens or a negatively refracting slab in the near-IR. On the other hand, ultra-highly doped semiconductors such as ZnO behave as 'mild' metal-like materials in the near-IR, which can enable high performance HMM-based devices. Furthermore, a material system such as ZnO and AZO has an evident advantage from the perspective of material technology: it allows epitaxial design of the device structure, which can further reduce losses and thus further boost the performance.

In our experiments, the planar HMM is formed by depositing 16 alternating layers of AZO and ZnO, each about 60 nm thick, on a silicon substrate. The deposition was carried out using laser ablation of $Al_2O_3$:ZnO (2 weight %) and ZnO targets. The deposition conditions of AZO were optimized previously to produce a low-loss plasmonic material[25]. The AZO/ZnO planar HMM was characterized using spectroscopic ellipsometer (V-VASE, J.A. Woollam Co.). The optical constants of the individual layers were extracted using a Drude + Lorentz model for AZO and a Lorentz model for ZnO (see Fig. 1a). In the near-IR range, AZO shows negative real permittivity for wavelengths longer than 1.84 μm, which allows AZO to be a metal substitute in the HMM design. Zinc oxide is a lossless dielectric with real permittivity close to 4 in the near-IR range. An alternating multilayer stack of AZO and ZnO with each elemental layer thickness being a small fraction (e.g., 1/10) of the wavelength can be approximated as a uniaxial medium in the effective-medium limit. The layered stack exhibits different permittivity values in the plane parallel to the layers and the direction perpendicular to the layers. The effective optical properties in these two directions were extracted from the ellipsometry data using a uniaxial model. Figure 1b shows the effective dielectric functions in the plane parallel to the layers ($\varepsilon_\parallel$) and in the direction perpendicular to the layers ($\varepsilon_\perp$). For wavelengths longer than 1.84 μm, the signs of the real permittivity components in the two different directions are opposite to each other, and therefore the material exhibits hyperbolic dispersion in this range. In the wavelength range from 1.84 μm to 2.4 μm, the multilayer stack exhibits Type-1 hyperbolic dispersion (Re$\{\varepsilon_\parallel\}$> 0 and Re$\{\varepsilon_\perp\}$< 0). Since Type-1 dispersion is essential for the phenomenon of negative refraction, we could expect this planar device to refract "negatively" in the wavelength range from 1.84 μm to 2.4 μm. Note that for wavelengths longer than 2.4 μm, the multilayer stack shows Type-2 hyperbolic dispersion.

The ellipsometry extraction of the optical parameters of the layered structure was verified by measuring far-field reflectance and transmittance in both transverse-magnetic (TM) and transverse-electric (TE) modes. The measurements were carried out at various angles of incidence ranging from 18 to 74 degrees. The measured reflectance spectra are plotted as color maps in the left panel of Fig. 2 for both incident polarizations. In comparison to the measured reflectance, the right panel of Fig. 2 plots the calculated reflectance spectra[26] for the multilayer stack. The calculations use material properties extracted from ellipsometer measurements as shown in Fig. 1a. The reflectance spectra from measurements and calculations are in good agreement, confirming the correctness of the ellipsometric extraction procedure for the optical parameters. The transmittance plots can be found in the supplementary information (see Fig. S2).

As mentioned previously, the multilayer stack of AZO/ZnO is expected to exhibit negative refraction in the wavelength range from 1.84 μm to 2.4 μm. In order to visualize this phenomenon, simulations were carried out using the extracted material parameters for AZO and ZnO. The field maps of the simulations in TM and TE polarizations are shown in Fig. 3. The TM-polarized Gaussian beam incident at an angle of 40 degrees to the sample normal undergoes negative refraction inside the sample, producing a beam shift towards the incident side of the sample normal. This dramatic effect is observed only in the case of TM-polarized (*p*-polarized) incident light because only in this polarization is the electric field component of the incident light able to probe both components of the dielectric function and thus experience the material's hyperbolic dispersion. In the case of TE-polarized (*s*-polarized) incident light, the medium behaves as an isotropic dielectric with a real permittivity given by $\varepsilon_\parallel$. Since observing the anomalous beam shift in the TM case would provide evidence for negative refraction, transmittance measurements were performed on the multilayer stack with half of the transmitted beam blocked by a razor blade. Figure 4a shows the schematic of the experimental set-up for observing negative refraction in our ZnO-based HMM. The position of the razor blade was adjusted to block the transmitted beam intensity exactly at 50% at normal incidence. We then increased the incidence angle and recorded the beam shift by measuring the relative transmittance (ratio of transmittance with and without the blade). It may be noted from Fig. 4a that the relative transmittance should dip if negative refraction were to occur. Figure 4b plots the ratio of the transmittance with and without the beam-blocking blade (referred to as relative transmittance). The figure also compares the measured curves with the theoretically predicted results. The theoretical predictions were carried out using the exact dispersion equation for a bulk, planar, multilayer stack[23] (see supplementary information for more details). The relative transmittance dips around 1.8 μm wavelength in the TM polarization due to negative refraction. No similar dip in the relative transmittance is observed for any incident angle in the TE-polarization (see supplementary information). The theoretical predictions agree well with the experiments, thus confirming our observation of negative refraction in the AZO/ZnO metamaterial.

The performance of HMMs can be characterized by a figure of merit (FoM). For a negatively refracting slab such as the device in this work, Hoffman *et al.* introduced the FoM as[15]: FoM = Re{$\beta_\perp$}/Im{$\beta_\perp$}, where $\beta_\perp$ is the propagation vector in the direction perpendicular to the layers. A large FoM requires smaller Im{$\beta_\perp$} (i.e., lower losses) and larger Re{$\beta_\perp$} (i.e., higher effective index). The FoM for our AZO/ZnO HMM was found to be about 12, which is at least three orders of magnitude larger than that of HMMs with a similar design using silver or gold[27]. In fact, the FoM in our sample is the highest FoM achieved so far in the optical range for any negatively refracting

HMM[4,21]. This conclusively shows that ultra-highly doped semiconductors can serve as good alternatives to metals for certain metamaterial applications in the near-IR.

In summary, we designed and fabricated a semiconductor-based hyperbolic metamaterial in the near-IR and in which AZO replaced conventional metals. Using the AZO/ZnO metamaterial, we successfully demonstrated negative refraction in the near-IR and showed that the figure of merit of this multilayered stack is about three orders of magnitude larger than that of its metal-based counterpart.

Furthermore, alternative plasmonic materials such as AZO overcome the bottleneck created by conventional metals in the design of optical metamaterials and enable more efficient devices. This demonstration opens up new approaches to engineer optical metamaterials and customize their optical properties in order to develop optical devices with enhanced performance. We anticipate that the development of new plasmonic materials and nanostructured material composites such as AZO/ZnO will lead to tremendous progress in the technology of optical metamaterials, enabling the full-scale development of this technology and uncovering many new physical phenomena.

**Acknowledgements**

This work was supported in part by the ONR grant N00014-10-1-0942 and the ARO grants W911NF-04-1-0350, 50342-PH-MUR and W911NF-11-1-0359.

**Author Contributions**

GN, VS and AB conceived the experiment idea, design and related discussions; GN performed the fabrication; GN and JL performed optical characterization; AK performed simulations; GN and AK performed theoretical calculations.

**Figure Captions**

**Figure 1**. Optical properties extracted from ellipsometry measurements. a) Dielectric functions of ZnO and Al-doped ZnO (AZO) in the near-IR. AZO is metallic for wavelengths longer than 1.84 µm. b) Dielectric functions of an alternating-layer stack of AZO and ZnO thin films (the inset shows the schematic of the layer stack) retrieved using a uniaxial model. The real and imaginary parts of permittivity in the plane parallel to the layers (represented as ∥) and perpendicular to the layers (represented as ⊥) are plotted in the near-IR. In the wavelength range from 1.84 µm to 2.4 µm, the layer stack shows Type-1 (Re{$\varepsilon_\parallel$} 0, Re{$\varepsilon_\perp$}< 0) hyperbolic dispersion. For longer wavelengths, the layer stack exhibits Type-2 (Re{$\varepsilon_\parallel$}< 0, Re{$\varepsilon_\perp$}> 0) hyperbolic dispersion.

**Figure 2**. The reflectance color maps of the layer stack sample in TM and TE polarizations for angles of incidence from 18 to 74 degrees. The left panels plot the measured reflectance spectra and the right panels plot the calculated reflectance spectra in both TM and TE polarizations.

**Figure 3.** Field maps showing the interaction of electromagnetic wave with AZO/ZnO multilayer stack. The incident Gaussian beam is TM polarized in the left panels and TE polarized in the right panels. The wavelength of the incident beam is 2 µm, and the angle of incidence is 40 degrees. The position of the AZO/ZnO multilayer stack is indicated by the orange rectangle. The bottom panels are the zoomed-in views of the top panels. The TM-polarization case exhibits negative refraction with respect to the sample normal and produces a significant beam shift to the incident side of the sample normal. In contrast with the TM case, the TE-polarized light is refracted positively and hence does not produce similar beam shift.

**Figure 4**. a) Schematic showing the experiemental set-up for the measurement of negative refraction. The transmitted beam is blocked partially by a razor blade before it reaches the detector. In the TM

case, the incident beam undergoes negative refraction and gets blocked more by the blade resulting in lower transmittance. In the TE case, the incident beam undergoes positive refraction and gets less blocked by the balde resulting in higher tranmittance. b) Relative transmittance (ratio of transmittance with and without a blade partially blocking the beam) for different angles of incidence from 0 to 50 degrees. The solid curves are the measured values, and the dashed curves are calculated from theory. The dip in the relative transmittance provides evidence of negative refraction in the case of TM polarization.

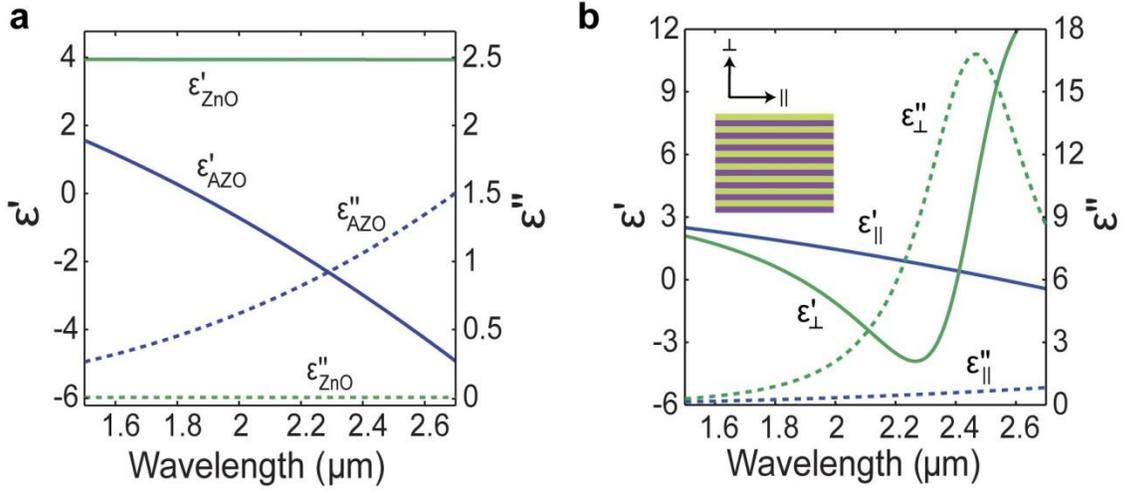

Figure 1

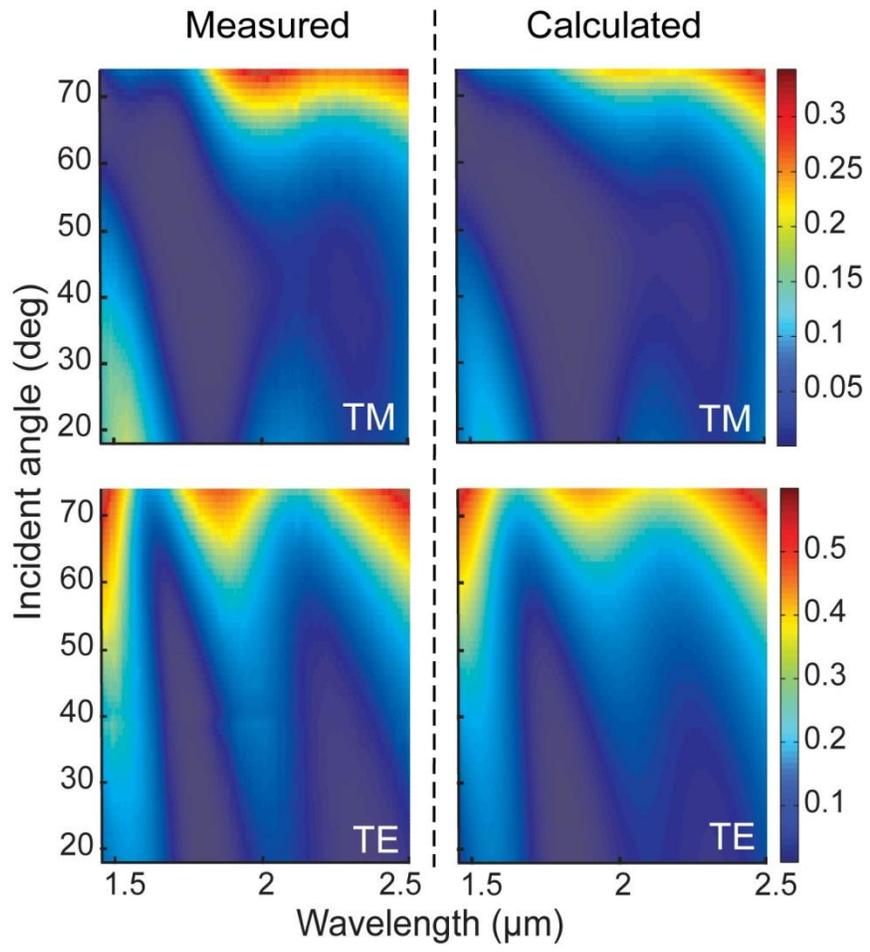

**Figure 2**

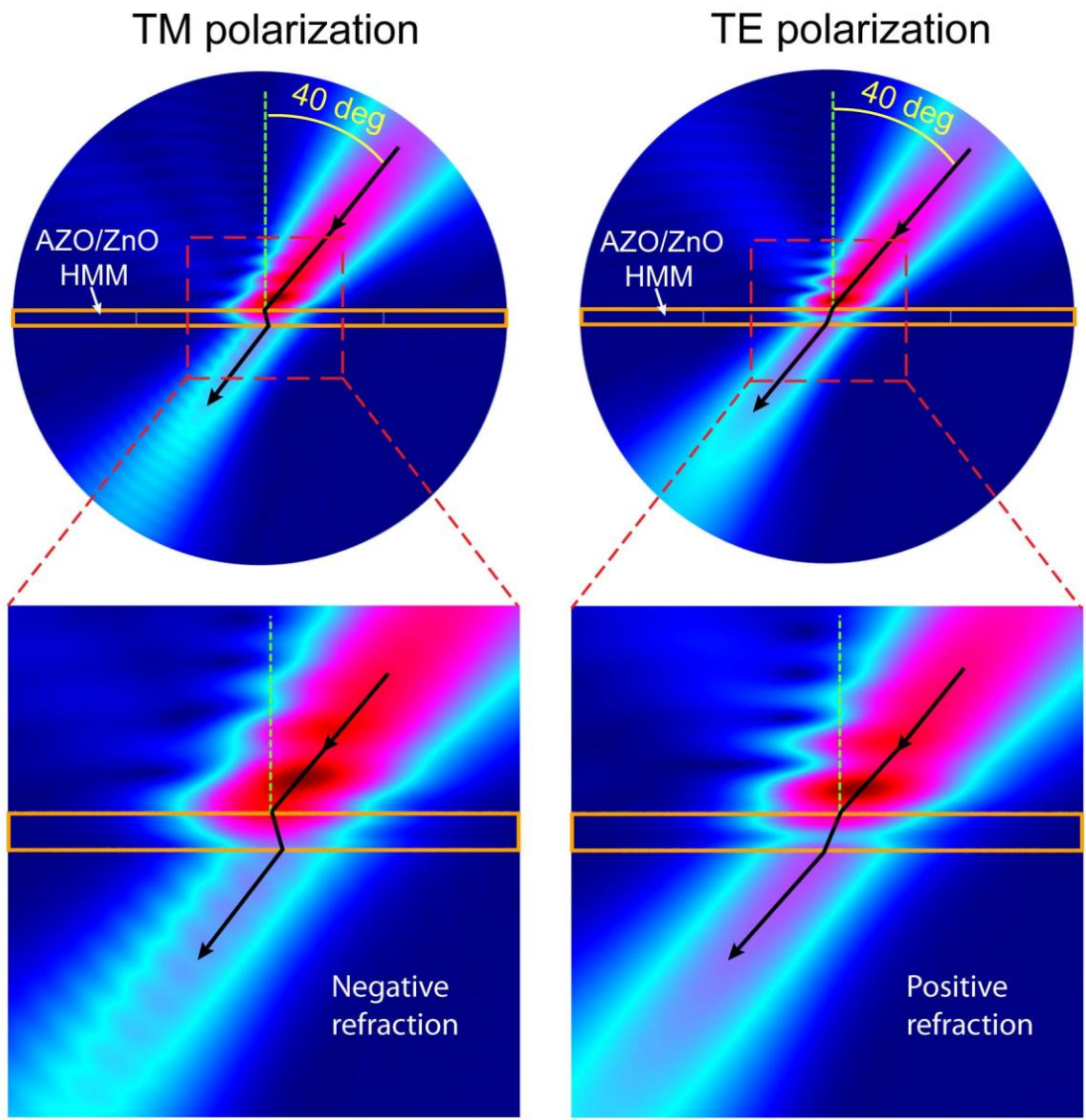

**Figure 3**

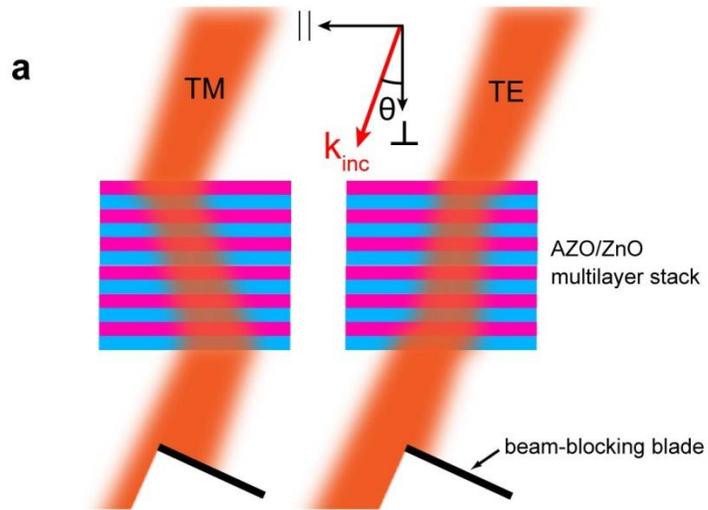

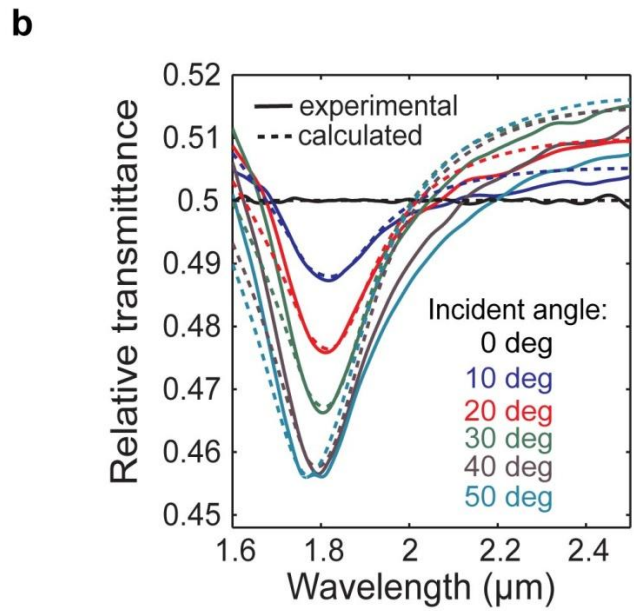

**Figure 4**

**Supplementary Information**

1. Effective parameters for the layer stack:

A medium arranged of alternating sub-wavelength layers of two materials (a binary lamellar medium) can be approximated by an anisotropic effective medium. If the dielectric functions of two elemental materials are $\varepsilon_m$ and $\varepsilon_d$ and the thicknesses of the layers are $d_m$ and $d_d$ respectively, the medium exhibits an uniaxial anisotropy with different permittivity values in the plane parallel to the layers (denoted as $\parallel$ direction) and the direction normal to the layers (denoted as $\perp$ direction).

The exact dispersion relations of electromagnetic waves in such a medium are described by equations (1) and (2). The wave vectors in the parallel ($\parallel$) and perpendicular ($\perp$) directions are denoted as $k_\parallel$ and $k_\perp$.

$$\cos[k_\perp(d_m + d_d)] = \cos(k_m d_m)\cos(k_d d_d) - \gamma \sin(k_m d_m)\sin(k_d d_d), \qquad (1)$$

where $\gamma$ is a polarization-specific parameter given by

$$\gamma_{TM} = \frac{1}{2}\left(\frac{\varepsilon_d k_m}{\varepsilon_m k_d} + \frac{\varepsilon_m k_d}{\varepsilon_d k_m}\right), \quad \gamma_{TE} = \frac{1}{2}\left(\frac{\varepsilon_d}{\varepsilon_m} + \frac{\varepsilon_m}{\varepsilon_d}\right), \qquad (2)$$

and $(k_q)^2 = \varepsilon_q(\omega/c)^2 - k_\parallel^2$, $q = \{m, d\}$.

Equation (1) is a non-linear equation (NLE), and the solutions to this equation for TM and TE modes provide effective permittivity values in the perpendicular and parallel directions, respectively. The permittivity values thus obtained are exact solutions for a bulk, infinite medium and show dependence on the incidence angle. Figure S1 plots the solutions of the NLE for an AZO/ZnO bilayer stack for three different angles (referenced to the sample normal). For comparison, the plot also includes the effective permittivity calculated from a first-order approximation of the NLE. Taking the first-order terms of the sine and cosine expansions and applying the small-angle approximation in equation 1, we arrive at an effective-medium expression for the first-order approximation, as shown in equation 3:

$$\varepsilon_\| = f\varepsilon_m + (1-f)\varepsilon_d, \quad \varepsilon_\perp^{-1} = f\varepsilon_m^{-1} + (1-f)\varepsilon_d^{-1}, \tag{3}$$

where $f = d_m/(d_m + d_d)$ is the metal volume filling fraction. We refer to this expression as the EMT-1 approximation.

It may be noticed from Fig. S1 that for larger angles, the EMT-1 approximation deviates significantly from the exact NLE solution. Hence, the NLE solutions are used to compute refraction angles, beam shift and relative transmittance shown in Fig. 4 in the paper.

2. Reflectance and transmittance of the layer stack:

The measured reflectance and transmittance spectra of the layer stack were compared to the calculated spectra to determine if the material parameters obtained from ellipsometry measurements were sufficiently reliable. The individual material properties obtained from ellipsometery (Fig. 1a) were used to compute the reflectance and transmittance spectra of the layer stack. The comparison of the reflectance spectra in TM and TE modes is presented in Fig. 2 in the paper. Figure S2 depicts the angular-dependent transmittance spectra of the sample. The measured and the calculated spectra agree reasonably well, confirming the correctness of the extraction of optical parameters from ellipsometry.

3. Confirmation of negative refraction:

The experimental set-up for observing negative refraction is shown in Fig. 4 in the paper. The measured data plotted in Fig. 4 was smoothed to eliminate noise and provide a better comparison between the measured and calculated curves. In Fig. S3, the raw data from the measurements are provided for both the TM and TE cases. The measured data are also compared against the calculated curves. The agreement is reasonable, confirming the observation of negative refraction in TM polarization only.

**Figure Captions:**

**Figure S1.** Effective permittivity values of the AZO/ZnO multilayer stack evaluated using a first-order effective medium theory (EMT-1) and the non-linear dispersion equation (NLE) as given in equation (1). The NLE solutions display angular dependence (angles are referenced to the surface normal direction). a) Real parts of permittivity ($\varepsilon'$), and b) Imaginary parts of permittivity ($\varepsilon''$) of the multilayer stack in the directions parallel ($\parallel$) and perpendicular ($\perp$) to layers.

**Figure S2.** Transmittance spectra of the AZO/ZnO multilayer stack in both the TM (upper panels) and TE (lower panels) polarizations for incident angles ranging from 18 to 74 degrees. The left panels plot the spectra measured and, the right panels plot the spectra obtained from theoretical calculations.

**Figure S3.** Measured raw data showing the relative transmittance (ratio of transmittances with and without a blade partially blocking the beam) of the AZO/ZnO layer stack for different incidence angles and polarizations. The dashed lines are the curves expected from theoretical calculations. The left panel plots relative transmittance for TM polarization and, the right panel plots it for TE polarization.

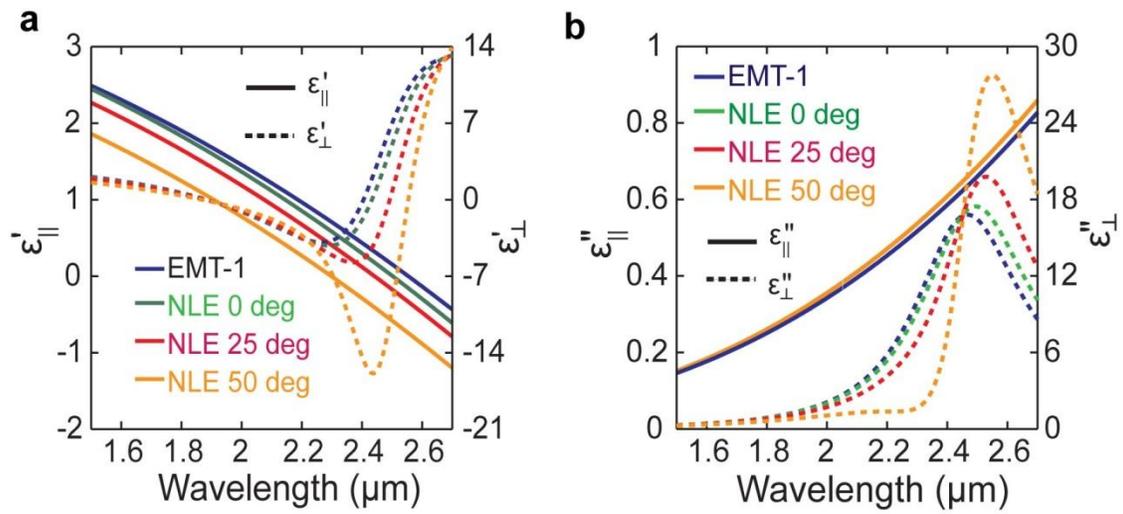

**Figure S1**

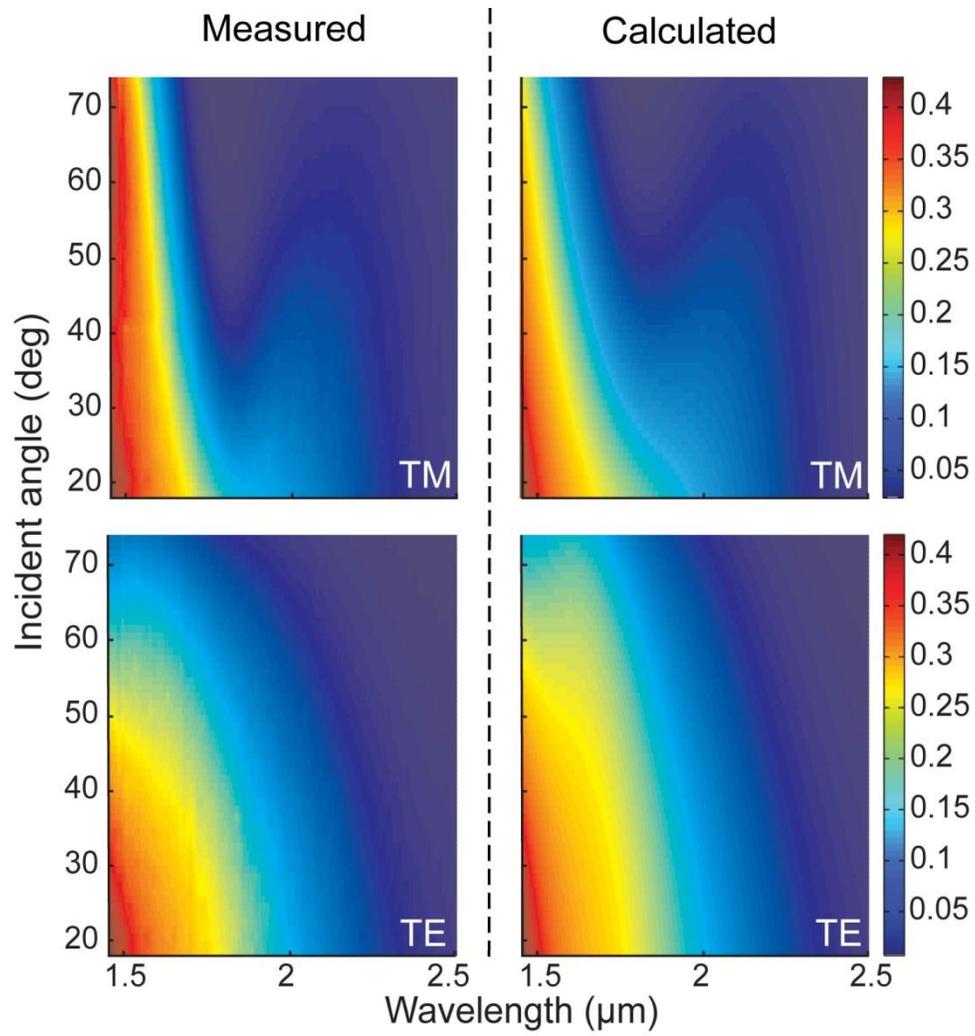

**Figure S2**

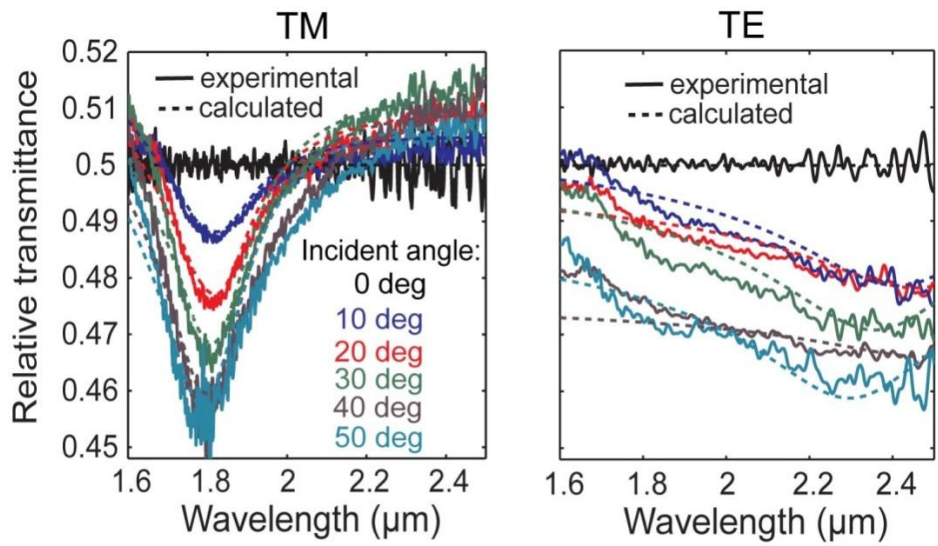

**Figure S3**